\documentclass[acus]{JAC2003}

\usepackage{graphicx}

\begin{document}
\title{A TEST FACILITY FOR THE INTERNATIONAL LINEAR COLLIDER \\
       AT SLAC END STATION A, \\
       FOR PROTOTYPES OF BEAM DELIVERY AND IR COMPONENTS\thanks{Work supported in part by U.S. Department of Energy contract DE-AC02-76SF00515, and by the Commission of the European Communities under the 6th Framework Programme "Structuring the European Research Area", contract number RIDS-011899.}} 
\author{ 
 M.~Woods,\thanks{mwoods@slac.stanford.edu}
 R.~Erickson,
 J.~Frisch, 
 C.~Hast,
 R.K.~Jobe,
 L.~Keller,
 T.~Markiewicz, \\
 T.~Maruyama, 
 D.~McCormick, 
 J.~Nelson, 
 T.~Nelson,
 N.~Phinney, 
 T.~Raubenheimer, \\ 
 M.~Ross, 
 A.~Seryi,
 S.~Smith, 
 Z.~Szalata, 
 P.~Tenenbaum, 
 M.~Woodley, 
 SLAC \\
 D.~Angal-Kalinin, 
 C.~Beard,
 C.~Densham,
 J.~Greenhalgh,
 F.~Jackson,
 A.~Kalinin, CCLRC \\
 F.~Zimmermann, CERN \\
 I.~Zagorodnov, DESY \\
 Y.~Sugimoto, KEK \\
 S.~Walston, LLNL \\
 J.~Smith, 
 A.~Sopczak, 
 D.~Burton, 
 R.~Tucker, 
 N.~Shales, Lancaster University \\
 R.~Barlow, 
 A.~Mercer, 
 G.~Kurevlev, Manchester University \\
 M.~Hildreth, Notre Dame University \\
 P.~Burrows,
 G.~Christian,
 C.~Clarke,
 A.~Hartin,
 S.~Molloy, 
 G.~White, QMUL, London \\
 W.~Mueller, 
 T.~Weiland, TEMF TU Darmstadt \\
 N.~Watson, University of Birmingham \\
 D.~Bailey,
 D.~Cussans, University of Bristol \\
 Y.~Kolomensky, University of California, Berkeley \\
 M.~Slater, 
 M.~Thomson, 
 D.~Ward, University of Cambridge \\
 S.~Boogert,
 A.~Liapine, 
 S.~Malton, 
 D.J.~Miller, 
 M.~Wing, UCL, London\\
 R.~Arnold, University of Massachusetts, Amherst \\
 N.~Sinev,
 E.~Torrence, University of Oregon \\
}

\maketitle

\begin{abstract}
The SLAC Linac can deliver damped bunches with ILC parameters for bunch charge and bunch length to End Station A. A 10Hz beam at 28.5 GeV energy can be delivered there, parasitic with PEP-II operation. We plan to use this facility to test prototype components of the Beam Delivery System and Interaction Region. We discuss our plans for this ILC Test Facility and preparations for carrying out experiments related to collimator wakefields and energy spectrometers. We also plan an interaction region mockup to investigate effects from backgrounds and beam-induced electromagnetic interference.
\end{abstract}

\section{INTRODUCTION AND OVERVIEW}

The International Linear Collider (ILC) is envisioned to be the next frontier accelerator facility for particle physics after the LHC begins operation, providing exceptional resolving power and precision for exploring the TeV energy scale.  New discoveries are expected in one or more exciting areas:  Higgs and the explanation of particle masses, supersymmetry, dark matter, extra dimensions, unification of fundamental forces.  The luminosity at this facility will be a factor 1000 greater than that achieved at LEP and 10,000 times greater than achieved at the SLC.  
      
In SLAC's End Station A (ESA), we are planning to test prototype components of the Beam Delivery System (BDS) and Interaction Region (IR).  This program~\cite{ESA} involves both machine and detector physicists, reflecting the close connections between the accelerator and experiment.  Primary areas of study are collimation, backgrounds and precision energy measurements.  The ESA experimental program includes many of the critical beam tests~\cite{criticaltests} discussed for the BDS at the 2004 KEK ILC Workshop.  It also plays an important role for the test beam program considered by the Worldwide Study on Detector test beams~\cite{worldtestbeam}.

\section{Beam Setup to ESA}
\begin{figure}[htb]
\centering
\includegraphics*[width=82.5mm]{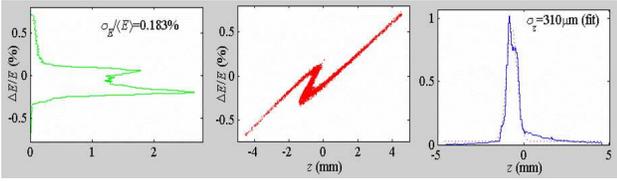}
%\vspace{-3mm}
\caption{Energy spread and Bunch length in ESA.}
\label{bunchparam}
\end{figure}

ESA beam tests are planned to run parasitically with PEP-II with single damped bunches at 10Hz, beam energy of 28.5 GeV and bunch charge of $2.0 \cdot 10^{10}$ electrons.  The long (5mm rms) bunch length out of the damping ring can be compressed in the Ring-to-Linac transfer line and in the 24.5-degree A-line bend from the Linac to ESA to achieve $\sim300 \mu$m bunch length in ESA.  A simulation with LiTrack~\cite{litrack} gives the results in Fig.~\ref{bunchparam} for the energy spread and bunch length in ESA.  Transverse beam sizes for the tests planned are expected to be $100-200 \mu$m rms.

\section{COLLIMATOR WAKEFIELDS}
At the ILC, collimators are required to remove halo particles (having large amplitudes relative to the ideal orbit) to minimize damage to beam line elements and particle detectors and to achieve tolerable background levels.  Short-range transverse wakefields excited by these collimators may perturb beam motion and lead to both emittance dilution and amplification of position jitter at the IP.

The goal of these tests~\cite{Txxx} is to find optimal materials and geometry for the collimator jaws to minimize wakefield effects while achieving the required performance for halo removal. The collimators will be rectangular in transverse section with a shallow longitudinal taper, long relative to the $\sim 300\mu m$ ILC bunch length.  To optimize their design, accurate modeling of wakefield effects for short bunches is needed, including the non-linear near-wall region which has implications for machine protection.  However, calculating the impedance for such an insertion using analytic methods is difficult, even for an idealized design without real engineering features such as contact fingers.  Similarly, tools such as MAFIA have problems in this regime due to grid dispersive effects. 

The ESA beam tests will allow further progress with both analytic calculations and development of state-of-the-art 3-d electromagnetic modeling methods~\cite{colwake-analytic}. Earlier measurements~\cite{colwake-recent} have already enabled significant development of analytic calculations, but the typical consistency with data is only within a factor 2-3.  The ILC design goal is agreement at the 10$\%$ level~\cite{colwake-TRC}.

\begin{figure}[htb]
\centering
\includegraphics*[width=70mm]{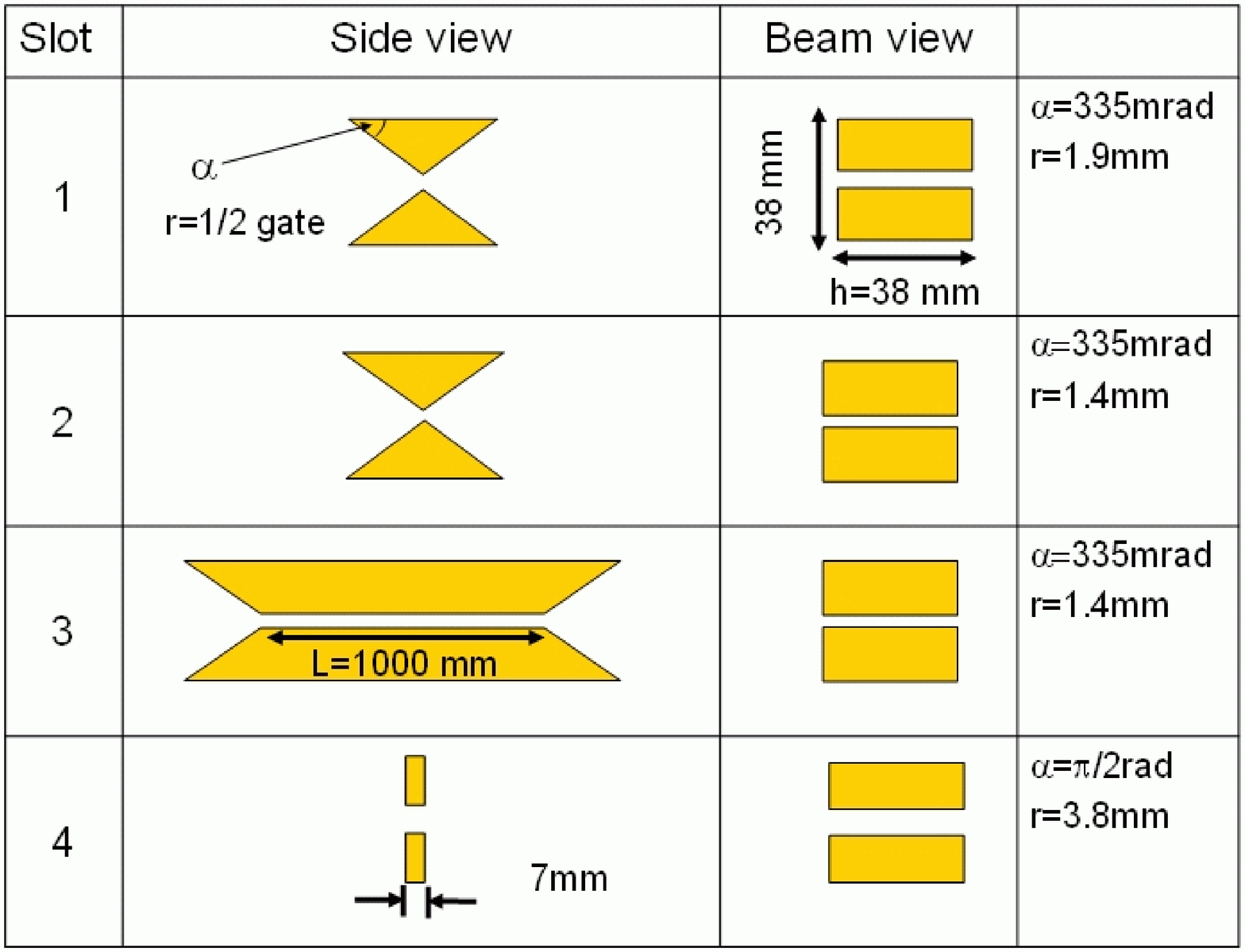}
%\vspace{-3mm}
\caption{Collimator insertions planned for a first set of measurements.}
\label{collim}
\end{figure}

Initial ESA measurements will measure resistive wakes in copper and study two-step tapers.  Two sets of four collimator insertions will be used, and Fig.~\ref{collim} shows the first set of four collimator insertions we plan to install in the Collimator Wakefield Box~\cite{colwake-box}.  Collimator 1 is the same as used in a recent measurement~\cite{colwake-recent}, and is essential for commissioning and control of systematics.  Two collimator sandwiches are available, each holding four collimator insertions.  We expect to be able to swap the sandwiches in an eight-hour shift.  Future plans could include investigations of different grades of copper, aluminium, and iron for resistive wakefields, and optimised tapers, as proposed in~\cite{colwake-analytic}.

\section{ENERGY SPECTROMETERS}
At the ILC, beam energy measurements with an accuracy of 100-200 parts per million (ppm) are needed for the determination of particle masses, including the top quark and Higgs boson.  Energy measurements both upstream and downstream of the collision point are foreseen by two different techniques to provide redundancy and reliability of the results~\cite{ipbiwhite}.  Upstream, a LEP-style beam position monitor (BPM) spectrometer is envisioned to measure the deflection of the beam through a dipole field.  Downstream of the IP, an SLC-style spectrometer is planned to detect stripes of synchrotron radiation (SR) produced as the beam passes through a string of dipole magnets.  

\begin{figure*}[tb]
\centering
\includegraphics*[width=100mm]{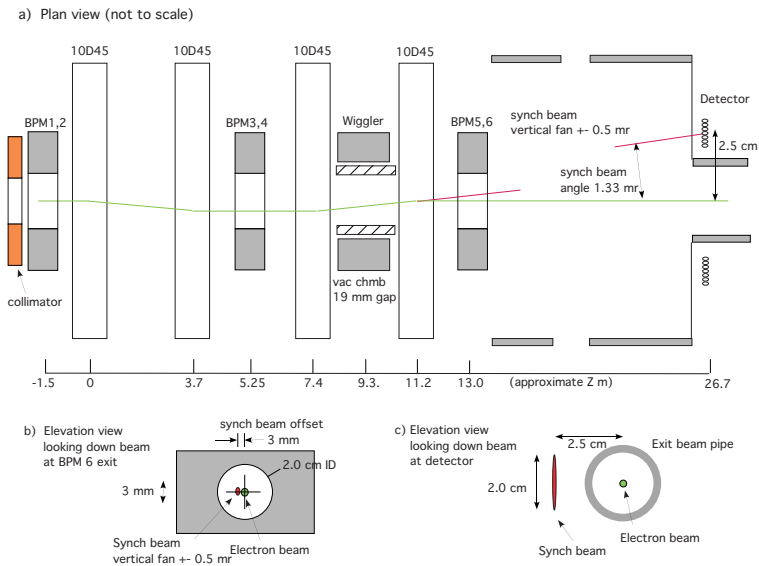}
%\vspace{-3mm}
\caption{Chicane for BPM and SR stripe energy spectrometer measurements.}
\label{energychicane}
\end{figure*}

In the proposed ESA tests, we plan to implement the BPM measurement and the synchrotron stripe technique in the same chicane (Fig.~\ref{energychicane}), which will have the same 5mm dispersion at mid-chicane and similar dipole fields ($\sim1$ kG) as the currently designed upstream ILC energy chicane.  It should be possible to measure beam offset with the BPM's at the same time as measuring the synchrotron light position, so beam energy determined from the two techniques can be compared directly.  Knowing the $\int B \cdot dL$ of the magnets and the chicane geometry, together with the position of the offset beam in the BPM's and the horizontal offset of the synchrotron light swath, the beam energy can be determined and compared with that delivered by the SLAC A-bend system.  To study systematics of the measurements we can dither the beam energy or trajectory at the end of the Linac, as well as change the beam trajectory in the ESA chicane.

\subsection{BPM energy spectrometer}
The BPM spectrometer tests~\cite{Txxx}, will use existing rf cavity BPMs (at least for the first stage), but will upgrade the BPM processing electronics.  The overall goal for system (mechanical and electrical) stability is $<\sim$500 nm, corresponding to 100 ppm energy precision over a one hour time scale.  One hour is a possible calibration timescale where one reverses polarity of the chicane and moves the BPMs at mid-chicane by 10mm on precision movers.  We are also investigating possibilities for a stretched wire or laser interferometer system to monitor mechanical stability of the support girder.  A 6-axis moveable stage would allow us to gauge the effects of beam tilts on the BPM measurements by observing any measurement bias as a function of overall BPM rotation.  We also plan temperature sensors along the girder to study temperature sensitivity of mechanical motion and electronics drifts.  

\subsection{SR Stripe Energy Spectrometer}
The SR stripe energy spectrometer~\cite{Txxx} must make precise measurements of the centroid and shape of the SR stripe.  A vertical SR stripe will be generated by a wiggler in the third leg of the chicane. (The ILC chicane will have wigglers in both the first and third legs; an additional wiggler in the first chicane leg in the ESA setup is a possible upgrade option.)  We are planning to test a detector array with 100$\mu m$ quartz fibers read out by a multi-anode PMT that senses Cherenkov light produced by secondary electrons generated by SR photon interactions in the fibers and the upstream window/radiator.  The beam test will validate Monte Carlo simulations of the Cherenkov light production and detection efficiency.  We will study backgrounds and cross-talk and compare energy measurements (energy jitter and absolute energy scale) with the BPM spectrometer and existing A-line diagnostics.  

\section{OTHER STUDIES:  IP BPMS, EMI}
We want to demonstrate that the fast IP BPMs and kickers, located within 4 meters of the IP, can work to the required precision in the presence of the intense beam-beam interaction.  We are studying simulating aspects of the beam-beam interaction with either a 5$\%$ radiation length fixed target or a "spray beam" to mimic a high flux of low energy pairs.  This system has been identified as one of the highest risks to delivering design luminosities for both the warm and cold LC designs, in part because of the difficulty of simulating the collision environment in a test beam.  We also plan to make measurements to quantify beam-induced electromagnetic interference (EMI) along the beamline and near features such as toroids, BPMs, and bellows.  We plan to measure the EMI frequency spectrum and its dependence on bunch charge and bunch length.

\end{document}